\documentclass{emulateapj}

\def\ga{\mathrel{\hbox{\rlap{\hbox{\lower4pt\hbox{$\sim$}}}\hbox{$>$}}}}
\def\la{\mathrel{\hbox{\rlap{\hbox{\lower4pt\hbox{$\sim$}}}\hbox{$<$}}}}

\usepackage{verbatim}
\usepackage{amsmath}
\usepackage{multirow}
\usepackage{relsize}
\usepackage{natbib}
\usepackage{amssymb}
\usepackage{float}

\newcommand{\sigs}{\sigma_s}

\newcommand{\meansigma}{\Sigma_0}

\newcommand{\means}{s_0}

\newcommand{\s}{\mathrm{s}}

\shorttitle{PDF Transition from Turbulence to Gravitational Instability}

\shortauthors{Burkhart, Stalpes,  \& Collins}
\begin{document}
\title{The Razor's Edge of Collapse: The Transition Point from Lognormal to Powerlaw in Molecular Cloud PDFs}
\author{Blakesley Burkhart\altaffilmark{1},  Kye Stalpes \altaffilmark{2}, \& David Collins \altaffilmark{2}}
\altaffiltext{1}{Harvard-Smithsonian Center for Astrophysics, 60 Garden st. Cambridge, Ma, USA}
\altaffiltext{2}{Department of Physics, Florida State University, Tallahassee, FL 32306-4350, USA}

\begin{abstract}

We derive an analytic expression for the transitional column density value ($s_t$) between the lognormal and power-law form of the probability distribution function (PDF) in star-forming molecular clouds.  
Our expression for $s_t$ depends on the mean column density, the variance of the lognormal portion of the PDF,
and the slope of the power-law portion of the PDF.  We show that $s_t$ can be related to physical quantities such as the sonic Mach number of the flow
and the power-law index for a self-gravitating isothermal sphere.  
This implies that the transition point between the lognormal and power-law density/column density PDF
represents the critical density where turbulent and thermal pressure balance, the so-called ``post-shock density.''
We test our analytic prediction for the transition column density using dust PDF observations reported in the literature
as well as numerical MHD simulations of self-gravitating supersonic turbulence with the Enzo code.
We find excellent agreement between the analytic $s_t$ and the measured values from the numerical simulations and observations (to within 1.5 A$_V$).
We discuss the utility of our expression for determining the properties of the PDF from unresolved low density material in dust observations, for estimating the post-shock density, and for determining the HI-H$_2$  transition in clouds.

\end{abstract}
\keywords{dust, extinction, galaxies: star formation, magnetohydrodynamics: MHD}

\section{Introduction}
\label{intro}

Star formation in galaxies occurs in dense molecular environments and is governed by the complex interaction of gravity, magnetic fields, 
turbulence, and radiation pressure \citep{McKee07,Elmegreen11}. 
Despite decades of study, the fundamental conditions behind the transition of diffuse atomic gas to cold molecular
gas are still relatively unconstrained \citep{Sternberg88,Krumholz09,MckeeKrum10,bialy15}. 
The initial conditions imprinted on the diffuse and molecular gas on parsec scales
(i.e. the level of turbulence, the cloud density, the structure of the magnetic field)
may determine the key properties of the initial mass function (IMF) and the star formation
rates in galaxies \citep{Hennebelle11}.   Therefore the properties of diffuse and molecular gas in and around
star-forming clouds must be quantified in order to construct a theory of star formation that predicts the IMF.

The density and column density probability distribution functions (PDFs) have been used extensively in understanding
the properties of galactic gas dynamics, from the diffuse ionized medium to dense star-forming clouds. 
The application of the PDF in molecular clouds has included density tracers such as CO \citep{lee12,burkhart13a}
and dust \citep{Kainulainen09,froebrich.rowles10, Schneider13b, Schneider14,schneider15,Lombardi15}.
Tracing the PDF using dust emission and absorption provides the 
largest dynamic range of densities, in contrast to molecular line tracers such as CO,
which suffer from depletion and opacity effects \citep{Goodman09,burkhart13a,burkhart13b}.

Simulations of self-gravitating MHD turbulence have successfully reproduced the shape and properties of the observational PDFs \citep{Burkhart09,Federrath12,Federrath13,Collins12,burkhart15a} suggesting
that the gas PDF stems from a combination of turbulence (which induces a lognormal PDF shape in density)
and self-gravity (which is characterized by a power-law PDF in density). In more detail, observed and simulated PDFs of Giant Molecular Cloud (GMC) environments, which include supersonic turbulence and self-gravity, suggest that 
the highest column density regime of the PDF (i.e., above column densities of 1 A$_V$) 
has a power-law distribution \citep{Collins11, Schneider13b,Lombardi15,Burkhart09,Federrath12,Federrath13,Collins12,burkhart15a}
while the lower column density material in the PDF is dominated by turbulent diffuse gas
and takes on a lognormal form \citep{Vazquez-Semadeni94,Burkhart12,Padoan97a}.

The implications for the shape of the gas density PDF in ISM clouds are profoundly linked
to the kinematics, star formation rates and the chemistry of the gas \citep{Federrath12}.
Kinematically, the PDF width of the lognormal density distribution can be related to the
sonic Mach number of the gas in an isothermal cloud \citep{Federrath08,Burkhart09,Kainulainen13,Burkhart12,burkhart15a}.
Star formation rates are linked to the gas density PDF in several analytic models  which use the
high density end of the PDF  to provide the dense
gas fraction to calculate star formation efficiencies \citep{Krumholz05,Hennebelle11,Padoan11}.
More recently, the HI PDF in and around GMCs has been proposed as a tracer of the HI-H$_2$ transition \citep{burkhart15, imara16}  as well as a more accurate tracer of the low density lognormal shape
as opposed to dust emission/absorption data, which have difficulty tracing the lognormal form \citep{Lombardi15,Schneider13b}.
\cite{burkhart15} and \cite{imara16} have shown that the lognormal portion of the column density PDF in a sample of Milky Way GMCs
is comprised of mostly atomic HI gas while the power-law tail is built up by the molecular H$_2$.
These studies suggest that the transition point  in the column density PDF between the lognormal
and power-law portions of the column density PDF traces important physical processes, such as the HI-H$_2$  transition and the density regime where self-gravity becomes dynamically important.  

In this work we derive an analytic formula for the transitional column density
from the lognormal portion of the PDF to the power-law form (denoted $s_t$).  
We organize the paper as follows. In Section 2 we derive an expression for the transitional column density for a  piecewise lognormal and power-law PDF distribution based on the assumption that the PDF is continuous and differentiable.  We further demonstrate that $s_t$ is related to the physical parameters such as the sonic Mach number of the gas (i.e. kinematics), the post-shock density, and the power-law index for a self-gravitating isothermal sphere.
In Section 3 we compare our analytic expression for the transitional column density to numerical simulations of self-gravitating MHD turbulence run using the Enzo code.
In Section 4 we compare our analytic expression for the transitional column density to observations using data from the literature.
In Section 5 we discuss our results, followed by our conclusions in Section 6.

\section{The Transition from  Lognormal to Power Law Tail in the PDF of a Turbulent Self-gravitating Medium}
\label{sec:ana}

The lognormal PDF of the gas column density is defined as
\begin{equation}
p_s(s)=\frac{1}{\sqrt{2\pi\sigs^2}}\exp\left(-\frac{(s-s_0)^2}{2\sigs^2}\right)\,,
\label{eq:pdf}
\end{equation}
with $s$ the logarithm of the normalized column density:
\begin{equation} \label{eq:s}
s\equiv\ln{(\Sigma/\meansigma)}\,.
\end{equation}

The PDF is a normal distribution in $s$, meaning that it is a lognormal distribution in $\Sigma$. The quantities $\meansigma$ and $\means$ denote, respectively, the mean column density and mean logarithmic column density, the latter of which can be related to the standard deviation $\sigs$ by:\footnote{This relationship was tested for a variety of molecular
clouds in \cite{Goodman09} and for MHD simulations in \cite{Price11}.}
\begin{equation}
\means=-\frac{1}{2}\sigs^2
\end{equation}

The lognormal form of the PDF of column density describes the behavior of diffuse HI and ionized gas \citep{berkhuijsen08,hill08, Burkhart10} as well as some star-forming molecular clouds that are not actively star-forming, e.g. see \cite{Kainulainen13,Schneider13b}.

The PDF of the highest column density regime of self-gravitating
turbulent clouds has a power-law distribution
as demonstrated in numerical simulations
\citep{Federrath12,Federrath13,Collins12,burkhart15a} and observations
\citep{Kainulainen09,froebrich.rowles10, Schneider13b,schneider15,Lombardi15,burkhart15}

Based on the aforementioned numerical and observational studies, hereafter we consider a piece-wise form for the PDF of column density (similar to the assumption of Collins et al. 2012 for the 3D density)
which has a lognormal distribution below a transitional column density value, denoted  $s_t=\ln (\Sigma_t/\Sigma_0)$, where $\Sigma_t$ is the
transitional column density value.
At column densities greater than $s_t$ the PDF is a power-law. We have

\begin{align}
p_s(s) = 
\begin{cases}
 N\frac{1}{\sqrt{2\pi}\sigs}\exp\left[-\frac{( s - s_0)^2}{2
 \sigs^2} \right ], & s < s_t \\
 N  p_0 \exp \left[{-\alpha s}\right ], & s > s_t ,
\end{cases}
\label{eqn.piecewise}
\end{align}
where again, $\means=-\frac{1}{2}\,\sigs^2$ and $p_0$ is the power-law's amplitude where it joins the lognormal.

Here the normalization $N$ is determined by the
normalization criterion, $\int_{-\infty}^{\infty} p_s(s) ds=1$, and is 
\begin{align}
N=   \left( p_0/\alpha e^{-\alpha s_t} +\frac{1}{2}\left[1+{\rm erf}\left( \frac{2 s_t
+\sigs^2}{2^{3/2} \sigs} \right) \right] \right)^{-1}
\end{align}

If we assume that $p_s(s)$ is continuous and differentiable, we can formulate an analytic
estimate for $\s_t$.  By setting the two parts of equation (4) equal at $s_t$ and setting their derivatives equal, we find
\begin{align}
s_t&=\frac{1}{2}(2|\alpha| -1)\sigs^2
\label{eqn.st}
\end{align}

The transition column density value between the lognormal and power-law PDFs therefore depends on the 
slope of the power-law tail ($\alpha$), the standard deviation of the lognormal
($\sigs$) and the mean column density, i.e. because $s_t=\ln(\Sigma_t/\Sigma_0)$ \footnote{We also solve for the power-law amplitude as:
$p_0=e^{\frac{1}{2}(\alpha-1)\alpha\sigma^2}/\sigma \sqrt{2\pi}  $}.
We note that the solution to the transition point should be applicable (and take the same form) for both density (see Collins et al. 2012) and column density
distributions since both density and column density share the same lognormal\footnote{The lognormal (Gaussian) form for column density is applicable under the condition that the
central limit theory can be applied, namely when the size of the emitting region is  larger than the decorrelation scale of turbulence  \citep{VazquezSemadeni2001}. } + power-law form of the PDF.
In the following subsection we provide a physical interpretation for $s_t$.

\subsection{Physical Interpretation of $s_t$}
\label{sec:ana}

The transitional column density  $s_t$ is not necessarily a criterion for a critical star-formation density, which most likely is farther out in the 
power-law tail.  Rather, $s_t$ represents a transitional point between the dominance of supersonic turbulence in the cloud gas dynamics, which builds the lognormal distribution,
to densities where gravity plays an increasingly important role in shaping the distribution.

Given the analytic solution for the transition point of the PDF between the lognormal and power-law
tail we are now in a position to relate the properties of the transition point to the physics of the 
gas in a GMC.  The width of the lognormal PDF ($\sigs$) depends on the properties of the turbulence in the GMC, with the primary dependence
being on the sonic Mach number.
For column density maps
\cite{Burkhart12} relate the sonic Mach number to PDF width as:

\begin{equation}
\sigs^2=A\ln[1+b^2M_s^2]
\label{eqn.sigma}
\end{equation}
$A=0.11$ is a scaling constant from density to column density.
The forcing parameter $b$ varies from
 $b\approx 1/3$ for purely solenoidal (divergence-free) forcing
to $b=1$ for purely compressive (curl-free) forcing of
MHD turbulence \citep{Federrath08}. 

Equation \ref{eqn.sigma} was shown to depend very weakly on the magnetic field \citep{Burkhart12}.  For the 3D density PDF in super-Alfv\'enic turbulence, \cite{Molina12} 
formulated the dependency on the plasma $\beta_0$, i.e. the ratio of the gas pressure to magnetic pressure, as

\begin{equation}
\sigma_{\ln\rho/\rho_0}^2=\ln[1+b^2M_s^2 \beta_0/(\beta_0+1)]
\label{eqn.beta}
\end{equation}

For the case of the column density, we can thus express the transition point in terms of the sonic Mach number by combining equations \ref{eqn.st} and \ref{eqn.sigma} to find
\begin{equation}
s_t=\frac{1}{2}(2|\alpha| -1)(A\ln[1+b^2M_s^2])
\label{eqn.rho0}
\end{equation}

\def\postshock{\rm{ps}}
The transition density or column density can be further expressed in terms of
the post-shock density,\footnote{This is referred to as the critical density in
\cite{Li15}.} $\rho_{\postshock}=\rho_0M_s^2$,
which is the density at which the turbulent energy density is equal to the thermal pressure:
\begin{equation}
P_{therm}=\rho_{\postshock}c_s^2=\rho_0v^2.
\end{equation}
Manipulating this relation we find that $M_s^2 = \rho_{\postshock}/\rho_0$, meaning equation \ref{eqn.rho0} becomes
\begin{equation}
s_t=(|\alpha| -1/2) A\ln[1+b^2\frac{\rho_{\postshock}}{\rho_0}]
\label{eqn.rho00}
\end{equation}

In the limit of strong collapse, $|\alpha|$ tends to 1.5 (see Figure
\ref{fig:SlopevTime}), so the $(|\alpha| -1/2)$ term is of order unity.

Therefore, 
\begin{equation}
\ln (\Sigma_t/\Sigma_0) \approx A\ln[1+b^2\frac{\rho_{\postshock}}{\rho_0}]
\label{eqn.rho1}
\end{equation}
and so
\begin{equation}
\Sigma_t/\Sigma_0 \approx (1+b^2\frac{\rho_{\postshock}}{\rho_0})^A.
\label{eqn.rho2}
\end{equation}

In the case of a 3D density field (relevant for simulations) we can express the transition density
in the same form at the column density (i.e. using Equation \ref{eqn.st})   the transition density can be expressed (using Equation \ref{eqn.beta}) as:

\begin{equation}
\rho_t/\rho_0 \approx (1+b^2\frac{\rho_{\postshock}}{\rho_0}\frac{\beta_0}{\beta_0+1})
\label{eqn.rhorho}
\end{equation}
as the exponent $A$ in Equation \ref{eqn.rho2} accounts for line-of-sight (LOS) effects and radiative transfer in column density (see Burkhart et al. 2013a).

The slope of the power-law tail, $\alpha$, does not have a clear relation to
other physical quantities.
It depends on the collapse state of the gas, the magnetic pressure, and the
LOS \citep{Kritsuk11,Ballesteros-Paredes11,Collins12,Federrath13,burkhart15a}.\footnote{We also note that the column density power-law slope $\alpha$ is related to, but not the same as, the power-law of the 3D density field.  The relation between these quantities is also derived in
\cite{Girichidis14}, their Equation 43.}  

In the case where the tail is produced only due to gravitational collapse, and if we assume
spherical symmetry, the PDF slope  of the power-law tail is related
to the exponent $\gamma$ of the radial density profile 
$\rho \sim  r^{-\gamma}$ \citep[e.g.][]{Shu77}.
Girichidis et al. (2014) showed analytically that column density power-law-tail slopes of $\alpha=-2.1$ 
correspond to the $\gamma=2$ prediction for a collapsing isothermal sphere since
$\alpha=-2/(\gamma - 1)$.

\section{Numerical Simulations}
\label{sec:num}

\subsection{Numerical Parameters and Methods}

We are now in a position to test the analytic relation for the transitional column density given in Equation \ref{eqn.st}.  For this purpose we use
simulation data generated by solving the ideal MHD equations including self-gravity using the AMR (Adaptive Mesh Refinement) code Enzo developed by
\cite{Collins10}.
These simulations use a root grid of $128^{3}$ with four levels of refinement to yield an effective resolution of $2048^{3}$. The Virial parameter $\alpha_{vir}$, sonic Mach number $\mathcal{M}_{s}$, and mean ratio of thermal to magnetic pressure $\beta_{0}$ are chosen here to be:  
$$\alpha_{vir} = 1$$ $$\mathcal{M}_{s} = 9$$ $$\beta_{0} = 0.2, 2.0, 20.0$$ which scale to physical clouds with free-fall time $t_{ff}$, box size $L_{0}$, rms velocity $v_{rms}$, total mass $M$ and mean magnetic field $B_{0}$ of:
$$t_{ff} = 1.1\;{\rm Myr}$$ $$L_{0} = 4.6\;{\rm pc}$$ $$v_{rms} = 1.8 \;{\rm km/s}$$ $$M = 5900\;M_{\odot}$$ $$B_{0} = (13, 4.4, 1.3) \mu G.$$
These simulations start with the same initial conditions as the simulations of
\citet{Collins12}, though they are down-sampled to the lower root grid resolution.  Also
the simulations of \citet{Collins12} were driven during the collapse, while
the present simulations were not.

These simulations have a post-shock density $\rho_{\postshock}/\rho_{0}=81$.  The
density may also be scaled physically using $\rho_{0} = 1000\ {\rm cm}^{-3}$ 
yielding a post-shock density of $\rho_{\postshock} = 8.1\times10{^4}\ {\rm cm}^{-3}$ 
or a column density of $\Sigma_{\postshock} =
6.7\times10^{23}\ {\rm cm}^{-2}$ given a cloud size of 4.6$\;{\rm pc}$.  Typical observational
values for the post-shock density range from $300\ {\rm cm}^{-3}$ to greater than
$4\times10^4\ {\rm cm}^{-3}$ \citep{Li15}.

\subsection{Column Density PDFs}

We project the 3D density into column density along three different lines of sight (denoted  $x, y,$ and $z$).
 A histogram is then generated of the logarithm of the normalized column density, i.e., $s=\ln(\Sigma / \Sigma_{0})$.  
Based on previous numerical studies (e.g., Collins et al. 2012; Burkhart et al.
2015) and the form of the PDF presented in Equation \ref{eqn.piecewise}, we
expect these simulations to show a lognormal column density PDF around the mean
column density with a power-law tail developing at higher densities. We
fit a lognormal to column densities within 20 percent of the peak of the
distribution (to minimize contamination from the tail) and a power-law in the higher density regions where the lognormal fit begins to fail.\footnote{Fits, analysis and plots are done at each time, magnetic field strength and line of sight noted above using the python packages yt, Simu, Scipy, Numpy and Matplotlib.}  

\begin{figure}[h]
\begin{center}
\includegraphics[width=0.49\textwidth]{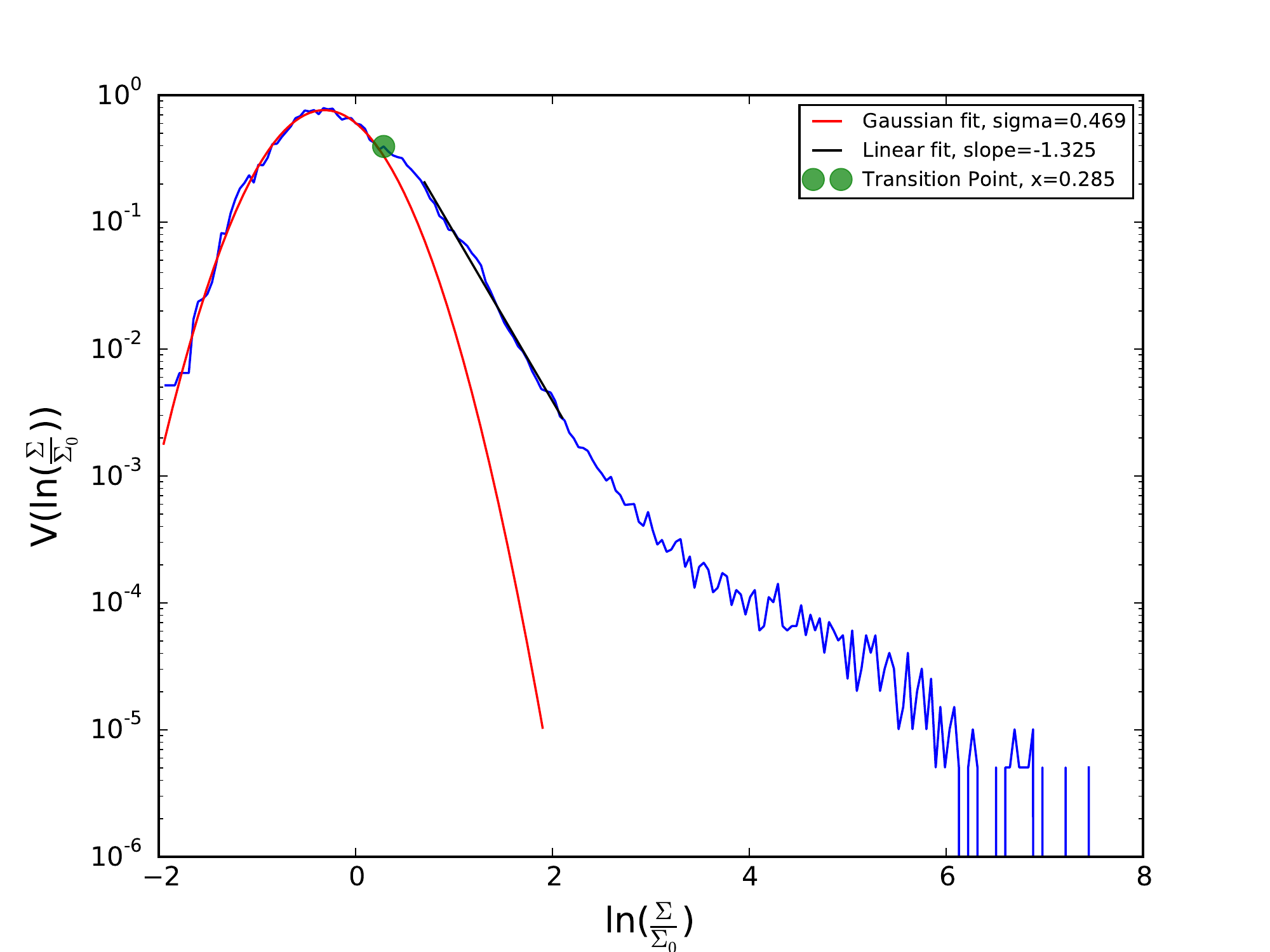}
\caption{An example PDF shown at $0.6t_{ff}$ for $\beta_{0} =$ 0.2  with line of sight along the y-axis. 
The transition point $s_{t}$ is indicated by the green dot, the lognormal fit is the red line, the power-law the black line and the actual data the blue line. 
}
\label{fig:pdfs}
\end{center}
\end{figure}

We show the PDF and the lognormal fits of the simulated column density at snapshot t=$0.6t_{ff}$ in Figure \ref{fig:pdfs}. The width of the lognormal $\sigma_{s}$ and power-law slope $\alpha$ are determined as free parameters of the fits while the transition point $s_{t}$ of the PDF is determined by finding where the least squares between the power-law fit and the data gets better than that of the lognormal fit. In Figure  \ref{fig:pdfs} the transition point $s_{t}$ is indicated by a green dot, the lognormal fit is a red line, the power-law a black line and the actual data a blue line.

\subsection{Numerical vs. Analytic Transitional Column Density}

We compare the analytic prediction for $s_t$ computed in Section 2  to the results found through fitting the simulation column density PDFs.

The value predicted for that width is approximately $0.5$ using the variables $\mathcal{M}_{s} = 9$, $b = 1/3$ and $A = 0.11$ as described in Section 2.

\begin{figure}[H]
\includegraphics[width=0.49\textwidth]{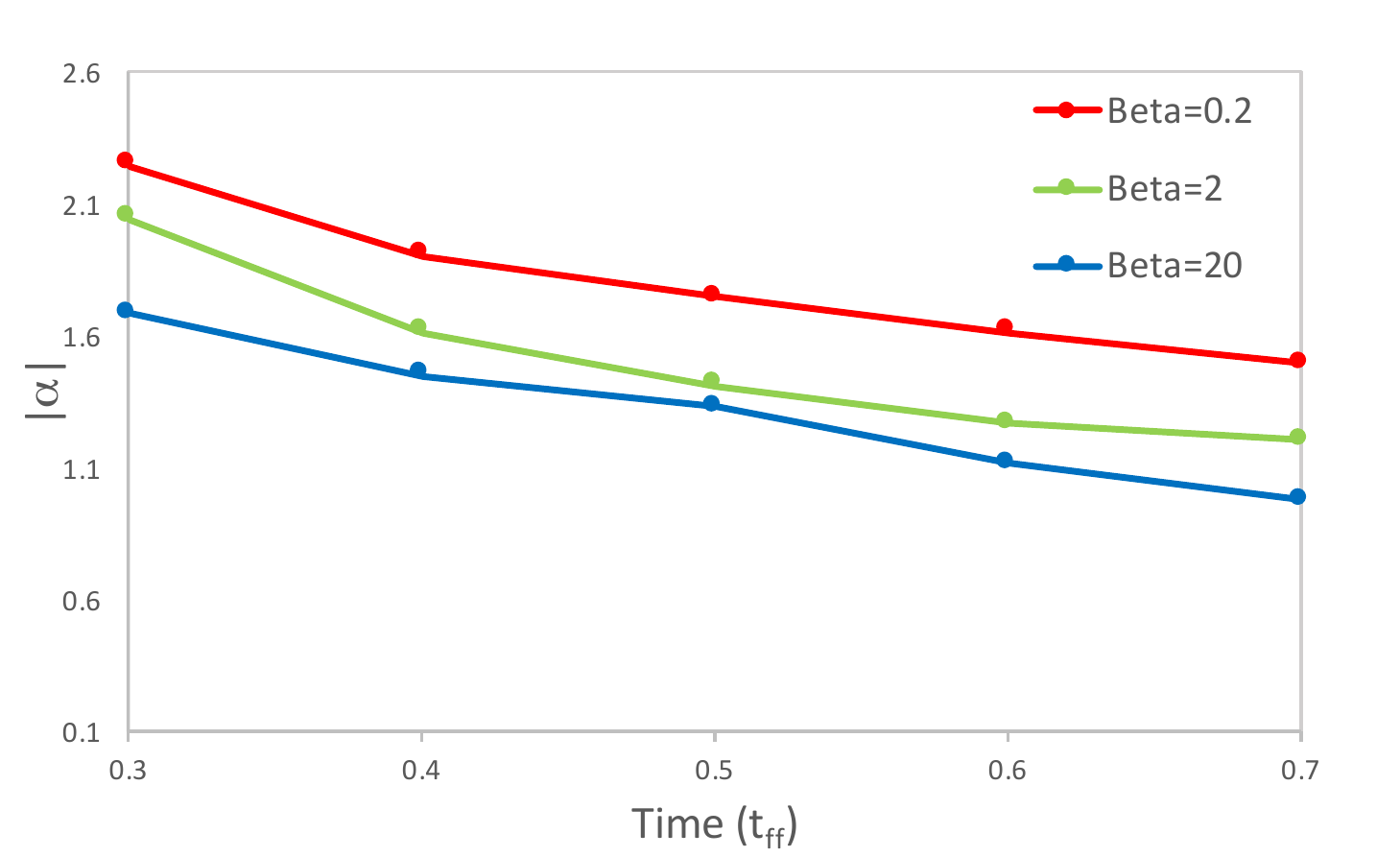}
\centering
\caption{Plot of $|\alpha|$ (y-axis) vs. $t$ (x-axis) for the range $0.3t_{ff}$ to $0.7t_{ff}$ where the power-law tail is well-developed.}
\label{fig:SlopevTime}
\end{figure}

\begin{figure}[H]
\includegraphics[width=\linewidth]{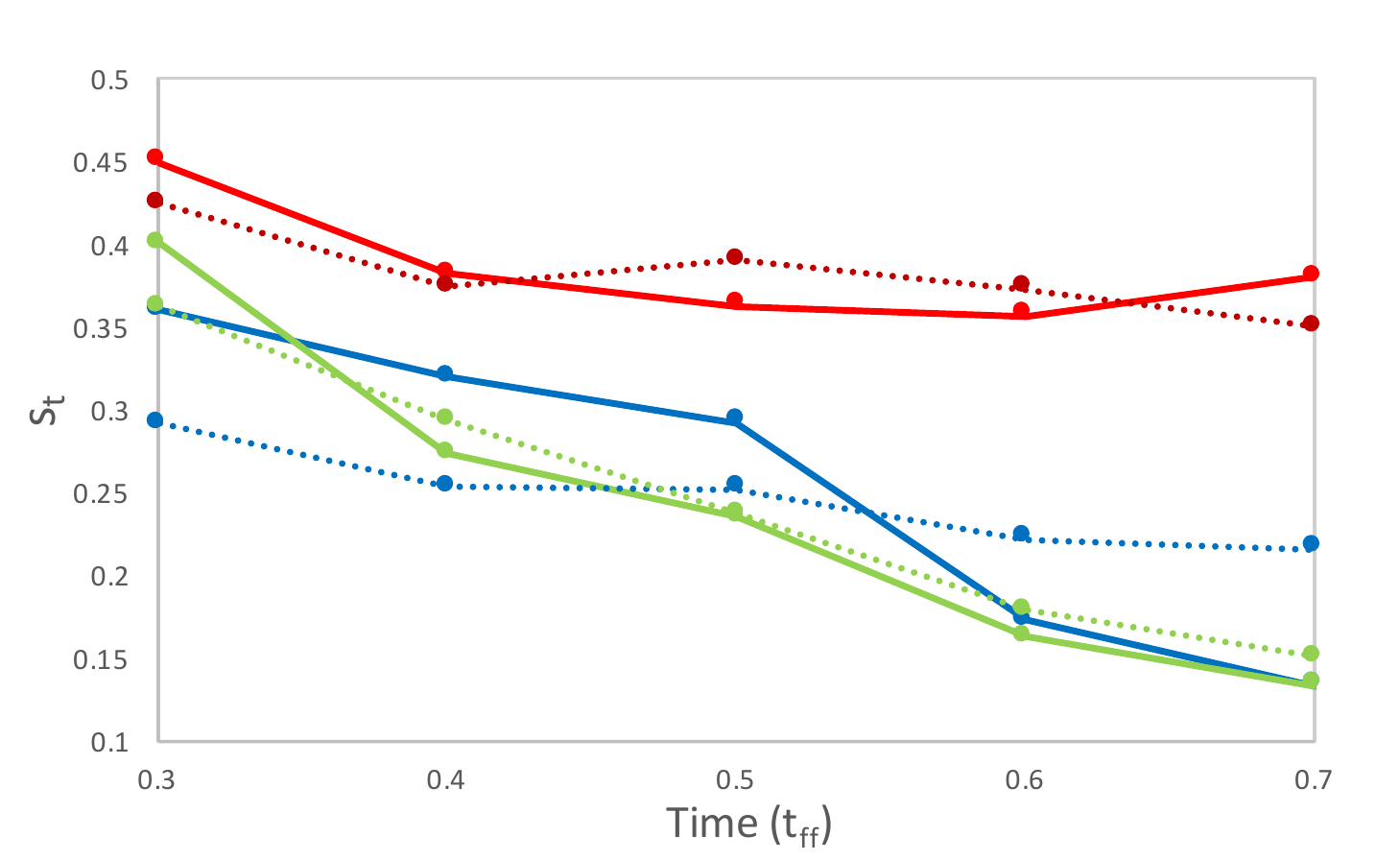}
\centering
\caption{Plot of transition point $s_{t}$ (y-axis) versus time (x-axis) for each magnetic field strength as predicted by equation 9 (dashed lines) and those found through fitting (solid lines) and colors as in Figure 2. }
\label{fig:Transitionpoint}
\end{figure}

Furthermore, the slope of the power-law tail is expected to decrease with time and with $\beta_{0}$,  as shown in Figure \ref{fig:SlopevTime}. 
The value of the power-law tail slope is roughly independent of the line of sight chosen (i.e. relative orientation to the mean magnetic field). 
Given the fitted width of the lognormal and the slope of the power-law tail, we compare the predicted value of the transitional column density $s_t$ to the measured value of the transitional column density, denoted $s_{t, fit}$. 
We present these results  in  Figure \ref{fig:Transitionpoint}.
We find excellent agreement between the predictions of the analytic fit proposed in Section 2 and the simulation results in Figure \ref{fig:Transitionpoint}.

\begin{figure}[h]
\includegraphics[width=\linewidth]{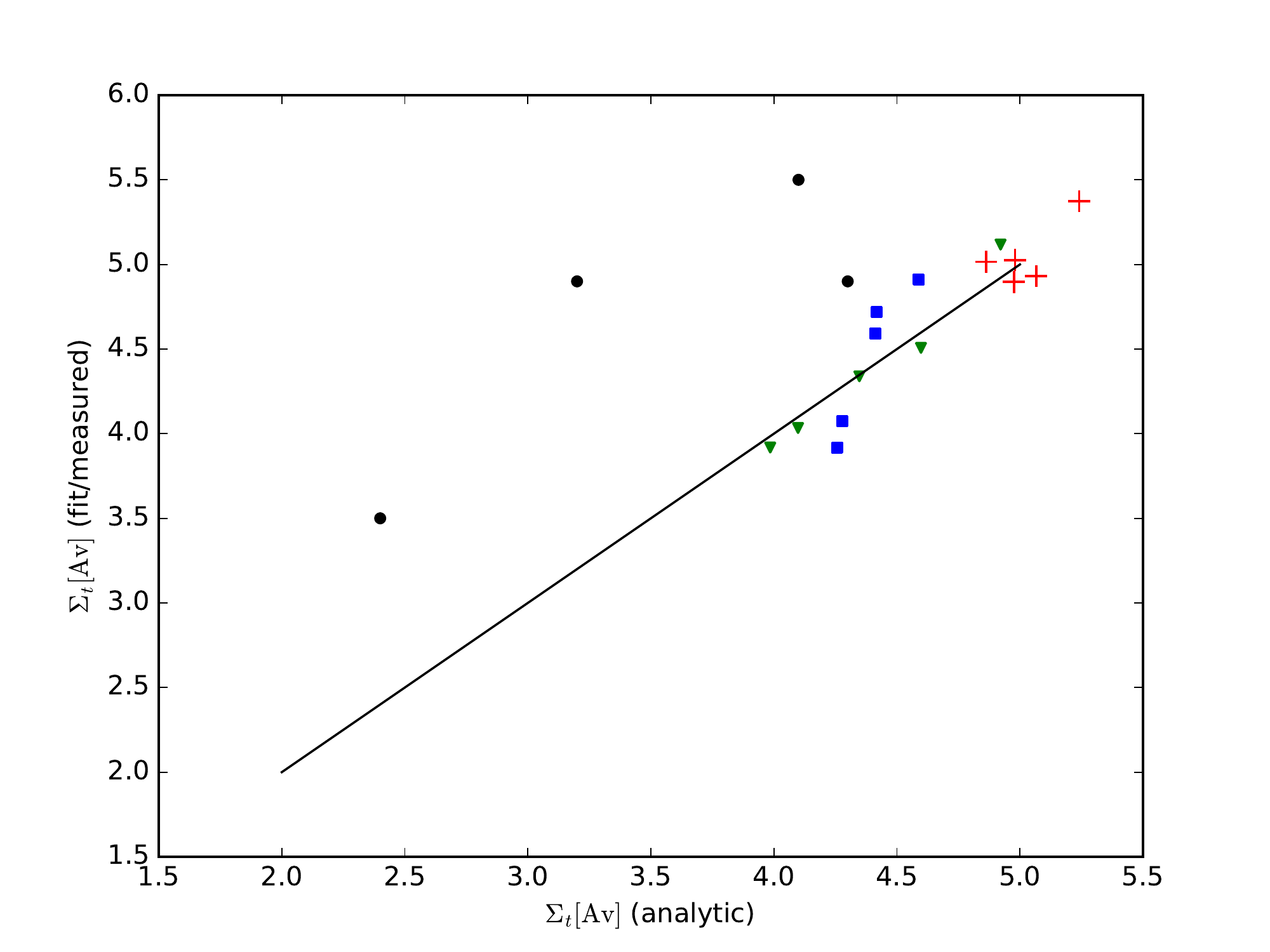}
\centering
\caption{Plot of fitted transition point vs analytic transition point for the three values of plasma $\beta$ along (colored points, blue square =$\beta=20$, green triangle=$\beta=2$, red +=$\beta=0.1$) with several observationally attained transition points (black circles) from nearby molecular clouds. }
\label{fig:Transition Points}
\end{figure}

\section{Observational Comparison}
In this section we test our analytic prediction for the transitional column density
against observations. In particular, Schneider et al. (2015, hereafter S15)
published values of the mean column density $\Sigma_0$, transitional column density ($\Sigma_t$), the width of the lognormal ($\sigma_s$), and the slope
of the power-law tail ($\alpha$) for four GMCs with different star formation histories 
and corrected for foreground and background dust contamination.  This provides
an observational test for comparing the predicted values of $\Sigma_t$ to the measured value, based on the measured values of $\Sigma_0$, $\alpha$ and $\sigma_s$
and the application of Equation \ref{eqn.st}.
We list the LOS foreground/background corrected parameters as reported in S15 and the analytic predicted value for  $\Sigma_t$ in Table \ref{tab:models}.

\begin{table*}
\begin{center}
\caption{Comparison of Measured and Predicted Transition Point.  Simulated
clouds are taken only at $t=0.5t_{\rm{ff}}.$
\label{tab:models}}
\begin{tabular}{ccccccc}
\hline\hline
Cloud & $\Sigma_0$ (Av) \footnote{Assuming N(H$_2$ ) = A$_v \times 0.94 \times
10^{21} {\rm cm}^{-2} /{\rm mag}^{-1}$} & $\alpha$& $\sigs$ & $\Sigma_{t,\rm{fit}}$ (Av)&
$\Sigma_{t,\rm{an}}$ (Av),    & reference \\
\hline\hline

NGC3603 & 3.4  &-1.31 & 0.52 & 4.9 &4.3 & S15 \\
Carina & 3.0 &-2.66  & 0.38  & 5.5 & 4.1&  S15 \\
Maddalena & 2.3   &-3.69 & 0.32 & 4.9 & 3.2& S15\\ 
Auriga & 1.6  &-2.54 & 0.45 & 3.5 & 2.4& S15 \\
$\beta=0.2$ & 3.4 & $\approx$ -1.5 & $\approx$ 0.5 & 4.9 & 5.1 & this work\\
$\beta=2$   & 3.4 & $\approx$ -1.5 & $\approx$ 0.5 & 4.3 & 4.3 & this work \\
$\beta=20$  & 3.4 & $\approx$ -1.5 & $\approx$ 0.5 & 4.6 & 4.4 & this work \\

\hline\hline
\end{tabular}
\end{center}
\end{table*}

The values of $\Sigma_{t, S15}$ and $\Sigma_{t, Eq. \ref{eqn.st}} $  agree to within approximately $1.5\; A_v$, with predicted values being consistently smaller.
We discuss the possible reasons for this in the next section.

\section{Discussion}
\label{sec:dis}

\subsection{The HI-H$_2$  Transition and Self-Gravity}

Recent studies have suggested that the PDF of molecular line tracers and dust tracers is of power-law form\citep{schneider15,Lombardi15} while the neutral diffuse 
HI builds up most of the lognormal portion of the PDF \citep{burkhart15,imara16}. 
In light of these recent studies, the  HI lognormal PDF and H$_2$ power-law tail PDF may  be effectively distinguished by the transition point between the two distributions.
The truncation of the HI lognormal roughly corresponds to the HI-H$_2$  transitional column density in Galactic star-forming clouds\citep{burkhart15,imara16} which suggests that measuring the transitional column density in such clouds could provide
constraints on the HI-H$_2$  transition. The transitional column density is approximately $\Sigma_t= 1-5\times 10^{21}\;{\rm cm}^{-2}$ (i.e. $\approx 8-38 \;M_\odot/{\rm pc}^2$) 
which is in the range of the typically quoted HI-H2 transition value of approximately $10\; M_\odot/{\rm pc}^2$\citep{MckeeKrum10}.
An example of this is recent observations of the Perseus molecular cloud, which find a HI-H$_2$  transitional column density of $\Sigma=9-11\times 10^{21}\;{\rm cm}^{-2}$

\subsection{Observational Properties of the Low Column Density PDF via $s_t$}

Recently several authors \citep{schneider15,Lombardi15} have noted that dust emission and extinction are problematic probes of the low column density material in molecular clouds. This is because the observed PDF of dust can suffer several biases including resolution, noise, boundary effects and  line-of-sight contamination.  
\cite{Lombardi15} pointed out that while the lognormal portion of the PDF cannot be securely traced by dust, the characteristic break in the
power-law regime at low values of extinction/column density  (i.e. $s_t$) is still unaffected by observational biases.  

These studies suggest that $s_t$ is a robust observational quantity, even though the properties of the lognormal PDF, such as the width of the lognormal, are not possible to accurately observe in dust tracers.\footnote{This is not true of other low column density tracers such as HI, which show characteristic lognormal distributions in column density and bimodal distributions in numerical simulations of density.}  Using our analytic expression for $s_t$ it is therefore possible to estimate the lognormal width of the distribution by measuring the power-law tail slope and value of s$_t$. The shape of the low-density portion of the PDF provides an important constraint on the initial conditions of star-forming clouds (i.e. the strength of turbulence and comparison to numerical studies) and therefore it is important to quantify this observationally. 

Incidentally, the difficulty of constraining the width of the PDF may be the reason that our predicted value for $s_t$ differs by about $1.5\; A_v$ from the Herschel observations reported in Table 1 (Schneider et al. 2015), since our prediction depends on the width of the PDF.   Since the measured values of $\alpha$ (slope of the power-law) and $s_t$ should be robust to observational effects, these two quantities should be used to measure the lognormal width $\sigma_s$, rather than fitting $\sigma_s$ directly from observations.

\section{Conclusions}
\label{sec:con}

The transition point between the turbulence-dominated (lognormal) portion of the PDF and the denser, self-gravitating (power-law) portion of the PDF is an important component of the star-formation process. 
In this paper we derived an analytic expression for the transitional point ($s_t$) of the column density PDF from a lognormal to 
a power-law.

We find that:
\begin{itemize}
\item The expression for $s_t$ depends on the mean column density, width of the lognormal portion of the PDF (i.e. the sonic Mach number and driving parameter)
and the slope of the power-law portion of the PDF (i.e. power-law index for a self-gravitating isothermal sphere)

\item  In the limit of strong collapse, $s_t$ represents the post-shock density given by the balance of turbulent and thermal pressure. 

 \item The values predicted by the analytic expression for $s_t$  agree well with measurements from Herschel dust observations and Enzo AMR simulations.
 
 \item The analytic expression reported in Equation \ref{eqn.st} will be useful for determining the properties of the PDF from unresolved low density material in observations
 and for estimating the HI-H$_2$  transition in clouds.
\end{itemize}

\acknowledgments
B.B. acknowledges support from the NASA Einstein Postdoctoral Fellowship. 
The authors are grateful to Shmuel Bialy, Zachary Slepian, and Amiel Sternberg for discussions on the meaning and derivation
of the transition point. This
work used the Extreme Science and Engineering Discovery Environment (XSEDE),
which is supported by National Science Foundation grant number ACI-1053575,
under allocation TG-AST140008.


\begin{thebibliography}{44}
\expandafter\ifx\csname natexlab\endcsname\relax\def\natexlab#1{#1}\fi

\bibitem[{{Ballesteros-Paredes} {et~al.}(2011){Ballesteros-Paredes},
  {V{\'a}zquez-Semadeni}, {Gazol}, {Hartmann}, {Heitsch}, \&
  {Col{\'{\i}}n}}]{Ballesteros-Paredes11}
{Ballesteros-Paredes}, J., {V{\'a}zquez-Semadeni}, E., {Gazol}, A., {Hartmann},
  L.~W., {Heitsch}, F., \& {Col{\'{\i}}n}, P. 2011, \mnras, 416, 1436

\bibitem[{{Berkhuijsen} \& {Fletcher}(2008)}]{berkhuijsen08}
{Berkhuijsen}, E.~M. \& {Fletcher}, A. 2008, \mnras, 390, L19

\bibitem[{{Bialy} {et~al.}(2015){Bialy}, {Sternberg}, {Lee}, {Le Petit}, \&
  {Roueff}}]{bialy15}
{Bialy}, S., {Sternberg}, A., {Lee}, M.-Y., {Le Petit}, F., \& {Roueff}, E.
  2015, \apj, 809, 122

\bibitem[{{Burkhart} {et~al.}(2015){Burkhart}, {Collins}, \&
  {Lazarian}}]{burkhart15a}
{Burkhart}, B., {Collins}, D.~C., \& {Lazarian}, A. 2015, \apj, 808, 48

\bibitem[{{Burkhart} {et~al.}(2009){Burkhart}, {Falceta-Gon{\c c}alves},
  {Kowal}, \& {Lazarian}}]{Burkhart09}
{Burkhart}, B., {Falceta-Gon{\c c}alves}, D., {Kowal}, G., \& {Lazarian}, A.
  2009, \apj, 693, 250

\bibitem[{{Burkhart} \& {Lazarian}(2012)}]{Burkhart12}
{Burkhart}, B. \& {Lazarian}, A. 2012, \apjl, 755, L19

\bibitem[{{Burkhart} {et~al.}(2013{\natexlab{a}}){Burkhart}, {Lazarian},
  {Ossenkopf}, \& {Stutzki}}]{burkhart13b}
{Burkhart}, B., {Lazarian}, A., {Ossenkopf}, V., \& {Stutzki}, J.
  2013{\natexlab{a}}, \apj, 771, 123

\bibitem[{Burkhart {et~al.}(2015)Burkhart, Lee, Murray, \&
  Stanimirović}]{burkhart15}
Burkhart, B., Lee, M.-Y., Murray, C.~E., \& Stanimirović, S. 2015, The
  Astrophysical Journal Letters, 811, L28

\bibitem[{{Burkhart} {et~al.}(2013{\natexlab{b}}){Burkhart}, {Ossenkopf},
  {Lazarian}, \& {Stutzki}}]{burkhart13a}
{Burkhart}, B., {Ossenkopf}, V., {Lazarian}, A., \& {Stutzki}, J.
  2013{\natexlab{b}}, \apj, 771, 122

\bibitem[{{Burkhart} {et~al.}(2010){Burkhart}, {Stanimirovi{\'c}}, {Lazarian},
  \& {Kowal}}]{Burkhart10}
{Burkhart}, B., {Stanimirovi{\'c}}, S., {Lazarian}, A., \& {Kowal}, G. 2010,
  \apj, 708, 1204

\bibitem[{{Collins} {et~al.}(2012){Collins}, {Kritsuk}, {Padoan}, {Li}, {Xu},
  {Ustyugov}, \& {Norman}}]{Collins12}
{Collins}, D.~C., {Kritsuk}, A.~G., {Padoan}, P., {Li}, H., {Xu}, H.,
  {Ustyugov}, S.~D., \& {Norman}, M.~L. 2012, \apj, 750, 13

\bibitem[{{Collins} {et~al.}(2011){Collins}, {Padoan}, {Norman}, \&
  {Xu}}]{Collins11}
{Collins}, D.~C., {Padoan}, P., {Norman}, M.~L., \& {Xu}, H. 2011, \apj, 731,
  59

\bibitem[{{Collins} {et~al.}(2010){Collins}, {Xu}, {Norman}, {Li}, \&
  {Li}}]{Collins10}
{Collins}, D.~C., {Xu}, H., {Norman}, M.~L., {Li}, H., \& {Li}, S. 2010, \apjs,
  186, 308

\bibitem[{{Elmegreen}(2011)}]{Elmegreen11}
{Elmegreen}, B.~G. 2011, \apj, 731, 61

\bibitem[{{Federrath} \& {Klessen}(2012)}]{Federrath12}
{Federrath}, C. \& {Klessen}, R.~S. 2012, ArXiv e-prints

\bibitem[{{Federrath} \& {Klessen}(2013)}]{Federrath13}
---. 2013, \apj, 763, 51

\bibitem[{{Federrath} {et~al.}(2008){Federrath}, {Klessen}, \&
  {Schmidt}}]{Federrath08}
{Federrath}, C., {Klessen}, R.~S., \& {Schmidt}, W. 2008, \apjl, 688, L79

\bibitem[{{Froebrich} \& {Rowles}(2010)}]{froebrich.rowles10}
{Froebrich}, D. \& {Rowles}, J. 2010, \mnras, 406, 1350

\bibitem[{{Girichidis} {et~al.}(2014){Girichidis}, {Konstandin}, {Whitworth},
  \& {Klessen}}]{Girichidis14}
{Girichidis}, P., {Konstandin}, L., {Whitworth}, A.~P., \& {Klessen}, R.~S.
  2014, \apj, 781, 91

\bibitem[{{Goodman} {et~al.}(2009){Goodman}, {Rosolowsky}, {Borkin}, {Foster},
  {Halle}, {Kauffmann}, \& {Pineda}}]{Goodman09}
{Goodman}, A.~A., {Rosolowsky}, E.~W., {Borkin}, M.~A., {Foster}, J.~B.,
  {Halle}, M., {Kauffmann}, J., \& {Pineda}, J.~E. 2009, \nat, 457, 63

\bibitem[{{Hennebelle} \& {Chabrier}(2011)}]{Hennebelle11}
{Hennebelle}, P. \& {Chabrier}, G. 2011, \apjl, 743, L29

\bibitem[{{Hill} {et~al.}(2008){Hill}, {Benjamin}, {Kowal}, {Reynolds},
  {Haffner}, \& {Lazarian}}]{hill08}
{Hill}, A.~S., {Benjamin}, R.~A., {Kowal}, G., {Reynolds}, R.~J., {Haffner},
  L.~M., \& {Lazarian}, A. 2008, \apj, 686, 363

\bibitem[{{Imara} \& {Burkhart}(2016)}]{imara16}
{Imara}, N. \& {Burkhart}, B. 2016, \apj, 771, 123

\bibitem[{{Kainulainen} {et~al.}(2009){Kainulainen}, {Beuther}, {Henning}, \&
  {Plume}}]{Kainulainen09}
{Kainulainen}, J., {Beuther}, H., {Henning}, T., \& {Plume}, R. 2009, \aap,
  508, L35

\bibitem[{{Kainulainen} \& {Tan}(2013)}]{Kainulainen13}
{Kainulainen}, J. \& {Tan}, J. 2013, \aap, 549, 53

\bibitem[{{Kritsuk} {et~al.}(2011){Kritsuk}, {Norman}, \& {Wagner}}]{Kritsuk11}
{Kritsuk}, A.~G., {Norman}, M.~L., \& {Wagner}, R. 2011, \apjl, 727, L20

\bibitem[{{Krumholz} \& {McKee}(2005)}]{Krumholz05}
{Krumholz}, M.~R. \& {McKee}, C.~F. 2005, \apj, 630, 250

\bibitem[{{Krumholz} {et~al.}(2009){Krumholz}, {McKee}, \&
  {Tumlinson}}]{Krumholz09}
{Krumholz}, M.~R., {McKee}, C.~F., \& {Tumlinson}, J. 2009, \apj, 693, 216

\bibitem[{{Lee} {et~al.}(2012){Lee}, {Stanimirovi{\'c}}, {Douglas}, {Knee}, {Di
  Francesco}, {Gibson}, {Begum}, {Grcevich}, {Heiles}, {Korpela}, {Leroy},
  {Peek}, {Pingel}, {Putman}, \& {Saul}}]{lee12}
{Lee}, M.-Y., {Stanimirovi{\'c}}, S., {Douglas}, K.~A., {Knee}, L.~B.~G., {Di
  Francesco}, J., {Gibson}, S.~J., {Begum}, A., {Grcevich}, J., {Heiles}, C.,
  {Korpela}, E.~J., {Leroy}, A.~K., {Peek}, J.~E.~G., {Pingel}, N.~M.,
  {Putman}, M.~E., \& {Saul}, D. 2012, \apj, 748, 75

\bibitem[{{Li} {et~al.}(2015){Li}, {McKee}, \& {Klein}}]{Li15}
{Li}, P.~S., {McKee}, C.~F., \& {Klein}, R.~I. 2015, \mnras, 452, 2500

\bibitem[{{Lombardi} {et~al.}(2015){Lombardi}, {Alves}, \& {Lada}}]{Lombardi15}
{Lombardi}, M., {Alves}, J., \& {Lada}, C.~J. 2015, \aap, 576, L1

\bibitem[{{McKee} \& {Krumholz}(2010)}]{MckeeKrum10}
{McKee}, C.~F. \& {Krumholz}, M.~R. 2010, \apj, 709, 308

\bibitem[{{McKee} \& {Ostriker}(2007)}]{McKee07}
{McKee}, C.~F. \& {Ostriker}, E.~C. 2007, \araa, 45, 565

\bibitem[{{Molina} {et~al.}(2012){Molina}, {Glover}, {Federrath}, \&
  {Klessen}}]{Molina12}
{Molina}, F.~Z., {Glover}, S.~C.~O., {Federrath}, C., \& {Klessen}, R.~S. 2012,
  \mnras, 423, 2680

\bibitem[{{Padoan} {et~al.}(1997){Padoan}, {Jones}, \& {Nordlund}}]{Padoan97a}
{Padoan}, P., {Jones}, B.~J.~T., \& {Nordlund}, A.~P. 1997, \apj, 474, 730

\bibitem[{{Padoan} \& {Nordlund}(2011)}]{Padoan11}
{Padoan}, P. \& {Nordlund}, {\AA}. 2011, \apj, 730, 40

\bibitem[{{Price} {et~al.}(2011){Price}, {Federrath}, \& {Brunt}}]{Price11}
{Price}, D.~J., {Federrath}, C., \& {Brunt}, C. 2011, in , 21

\bibitem[{{Schneider} {et~al.}(2013){Schneider}, {Andr{\'e}}, {K{\"o}nyves},
  {Bontemps}, {Motte}, {Federrath}, {Ward-Thompson}, {Arzoumanian},
  {Benedettini}, {Bressert}, {Didelon}, {Di Francesco}, {Griffin}, {Hennemann},
  {Hill}, {Palmeirim}, {Pezzuto}, {Peretto}, {Roy}, {Rygl}, {Spinoglio}, \&
  {White}}]{Schneider13b}
{Schneider}, N., {Andr{\'e}}, P., {K{\"o}nyves}, V., {Bontemps}, S., {Motte},
  F., {Federrath}, C., {Ward-Thompson}, D., {Arzoumanian}, D., {Benedettini},
  M., {Bressert}, E., {Didelon}, P., {Di Francesco}, J., {Griffin}, M.,
  {Hennemann}, M., {Hill}, T., {Palmeirim}, P., {Pezzuto}, S., {Peretto}, N.,
  {Roy}, A., {Rygl}, K.~L.~J., {Spinoglio}, L., \& {White}, G. 2013, \apjl,
  766, L17

\bibitem[{{Schneider} {et~al.}(2014){Schneider}, {Ossenkopf}, {Csengeri},
  {Klessen}, {Federrath}, {Tremblin}, {Girichidis}, {Bontemps}, \&
  {Andre}}]{Schneider14}
{Schneider}, N., {Ossenkopf}, V., {Csengeri}, T., {Klessen}, R., {Federrath},
  C., {Tremblin}, P., {Girichidis}, P., {Bontemps}, S., \& {Andre}, P. 2014,
  ArXiv e-prints

\bibitem[{{Schneider} {et~al.}(2015){Schneider}, {Ossenkopf}, {Csengeri},
  {Klessen}, {Federrath}, {Tremblin}, {Girichidis}, {Bontemps}, \&
  {Andr{\'e}}}]{schneider15}
{Schneider}, N., {Ossenkopf}, V., {Csengeri}, T., {Klessen}, R.~S.,
  {Federrath}, C., {Tremblin}, P., {Girichidis}, P., {Bontemps}, S., \&
  {Andr{\'e}}, P. 2015, \aap, 575, A79

\bibitem[{{Shu}(1977)}]{Shu77}
{Shu}, F.~H. 1977, \apj, 214, 488

\bibitem[{Sternberg(1988)}]{Sternberg88}
Sternberg, A. 1988, ApJ, 332, 400

\bibitem[{{Vazquez-Semadeni}(1994)}]{Vazquez-Semadeni94}
{Vazquez-Semadeni}, E. 1994, \apj, 423, 681

\bibitem[{{Vazquez-Semadeni} \& {Garcia}(2001)}]{VazquezSemadeni2001}
{Vazquez-Semadeni}, E. \& {Garcia}, E. 2001, ApJ, 557, 727

\end{thebibliography}
\end{document}